\documentclass[usenatbib,a4paper]{mn2e}
\usepackage[psamsfonts]{amssymb}
\usepackage{graphics}
\def\mdot{{\dot{M}}}

\def\msun{{\rm M_{\odot}}}

\def\eg{{e.g.}}

\def\etal{{et al.}}

\def\pie{{ \pi}}

\def\bphi{B_{{\varphi}}}
\def\bz{B_{{z}}}
\def\omegak{\Omega_{\rm{k}}}
\def\omegastar{\Omega_{\rm{\star}}}

\def\rc{r_{\rm{c}}}
\def\amag{a_{\rm{mag}}}

\begin{document}
\title[The fate of short period cataclysmic variables]
{Propeller activated resonances and the fate of short period cataclysmic variables}
\author[O.M. Matthews et al.]{O.M. Matthews$^1$\footnotemark[1], P.J. Wheatley$^2$, G.A. Wynn$^3$ and M.R. Truss$^4$\\
$^1$ Laboratory for Astrophysics, Paul Scherrer Institut, W\"{u}renlingen und Villigen, CH-5232 Villigen PSI, Switzerland\\
$^2$ Department of Physics, University of Warwick, Coventry, CV4 7AL\\
$^3$ Department of Physics \& Astronomy, University of Leicester, University Road, Leicester, LE1 7RH \\
%$^4$ School of Physics \& Astronomy, University of St Andrews, North Haugh, St Andrews, Fife, KY16 9SS
$^4$ Department of Physics, Durham University, South Road, Durham, DH1 3LE
} 

%\date{Received ** *** 2003; in original form 2003 *** **}

\label{firstpage}

\maketitle

\begin{abstract}
We show that the combination of a weak 
magnetic propeller and accretion disc resonances can effectively halt accretion in short period 
cataclysmic variables for large fractions of their lifetimes. This may 
help to explain the discrepancy between the observed and 
predicted orbital period distributions of cataclysmic variables at short 
periods. 
Orbital resonances cause the disc to become eccentric, allowing 
material to fall back onto the donor star or out of the system. A weak 
magnetic field on a rapidly spinning primary star propels disc material 
outwards, allowing it to access these resonances. Numerical and analytic 
calculations show 
that this state can be long lived ($\sim10^{11}$\,yr).
This is because the magnetic 
propeller is required only to maintain access to the resonances, and not to push matter out of the Roche lobe, so that the
spin down time-scale is much longer than for a classical propeller model.

\end{abstract}

\begin{keywords}
% Keywords selected from
% http://www.blackwell-science.com/~cgilib/jnlpage.asp?Journal=mnras&File=mnras&Page=authors/keywords (maximum of six)
accretion, accretion discs  --  stars: magnetic fields  --  stars: dwarf novae
\end{keywords}

\footnotetext[1]{E-mail: owen.matthews@psi.ch}

\section{Introduction}
\label{sec:int}

It is known that accretion disc resonances in cataclysmic variables can 
remove angular momentum from the outer accretion disc \citep{whi88}. This occurs when an element of the disc orbits with a frequency which is resonant with that of the binary orbit, typically at a ratio of three to one. This 3:1 resonance drives the accretion disc to become eccentric and causes the superhump modulations which are the defining feature of the superoutbursts of cataclysmic variables. Disc resonances provide a second source of angular momentum transport which is responsible for increasing the mass transfer rate, driving ordinary outbursts to become superoutbursts \citep*{osa89,tru01}.

In known systems, superhumps normally fade soon after outburst. 
This is because the tidal torques clear material from the outer disc until there is no material remaining at the resonant radius. Resonant 
effects cannot switch on again until a new outburst 
drives the outer disc beyond the resonant radius once more.

In this paper we consider the possibility that access to orbital resonances 
can be maintained almost indefinitely by the action of a weak magnetic propeller. 
It is known that
rapidly spinning and highly magnetic primary stars can act as magnetic 
propellers in some cataclysmic variables. Indeed, in the case of AE~Aqr it is believed 
that the entire mass transfer stream is ejected by a particularly strong 
magnetic propeller \citep*{wyn97} and the evolution of AE~Aqr \citep*{sch04} and WZ~Sge \citep*{las99} may well have been influenced by magnetic effects. Here we consider the possibility that a 
{\it much weaker} magnetic propeller might drive the disc outwards as far as the resonant radius, and so 
maintain a reservoir of material at this radius. The resulting 
continuous deposition of angular momentum by the resonance would drive the 
growth of eccentric modes. Disc material would be therefore be likely to extend beyond the primary Roche lobe. Such material
would then either escape the binary or fall back to the donor star, preventing accretion onto the primary star in either case.  

We are motivated to study this mechanism because it may 
modify the evolution of cataclysmic variables, especially if this state 
can be long lived. This may well be the case, since a much weaker torque is required than in the AE~Aqr case. The mechanism may slow or even halt accretion onto the white dwarf by driving mass out of the primary Roche lobe. 
At smaller mass ratios the resonant radii move further into the Roche lobe of the primary star, and are more likely to reach the disc,
so that any effect is likely to influence short period cataclysmic variables 
disproportionately. This is interesting because existing 
models are unsuccessful in reproducing the observed evolutionary behaviour 
of short period cataclysmic variables. 

\begin{figure}
\begin{center}
\resizebox{80.0mm}{80.0mm}{
\mbox{
\includegraphics{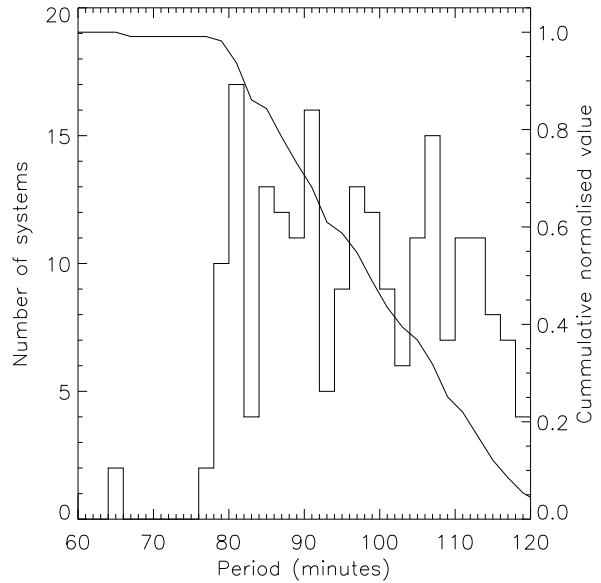}}}
\end{center}
\caption{CV orbital period histogram with the cumulative frequency overplotted as a solid line. This plot is based on fig. 1 from \citet{kin02}, but uses more recent data from then online version of \citet{rit98}. The plot contains data for all 215 CVs with periods known to be between $60$ and $120 \ {\rm{ mins}}$, from a total of 569 with known periods in the catalogue.}
\label{fig:klaus}
\end{figure}

Binary evolution models predict a minimum orbital period for cataclysmic 
variables which is reached when the donor star is no longer able to shrink in 
response to mass loss. Models consistently place this minimum at 
around 65\,min \citep*[e.g.]{kol92,how97,kol99}, whereas the observed period distribution shows a sharp cut off at 78\,min \citep*[e.g.][]{rit98,tap02}. The same models also predict that the evolution of the orbital period should slow at short periods, leading to an accumulation of systems in a sharp {\it period spike} at the minimum period. This feature is absent from the observed distribution, as shown in Fig.~\ref{fig:klaus}.

A number of explanations have been proposed for these discrepancies, but none 
are entirely satisfactory. An obvious solution is to hide short period systems 
from view, but standard selection effects do not act sufficiently strongly to explain the differences \citep*{pat98,kin02}. 
Another possibility is that an additional angular momentum loss mechanism 
could act to increase the minimum period \citep{pat98} or a range of 
angular momentum loss rates could blunt the period spike \citep{kin02}. However, \citet{bar02}
show that even the loss of all the angular momentum carried by the accretion 
stream cannot increase the minimum period sufficiently. They also argue that blunting the period spike by `smearing'  the angular momentum loss rates would require a great deal of fine tuning to be successful.

In this paper we consider how the effects of propeller-induced access to orbital resonances might prevent some short-period CVs from exhibiting outburst behaviour, and cause them to become fainter. In Section~\ref{sec:bin} we discuss the model in more detail. In Sections~\ref{sec:num} and \ref{sec:res} we explore the parameter space for which the idea may be of interest using smoothed particle hydrodynamics calculations. Finally, in Section~\ref{sec:dis}, we discuss how this mechanism might influence the evolution of short period cataclysmic variables.
%In Section\,\ref{sec:ana} we test analytically whether this state can be long lived.

\begin{figure}
\begin{center}
\resizebox{80.0mm}{80.0mm}{
\mbox{
\includegraphics{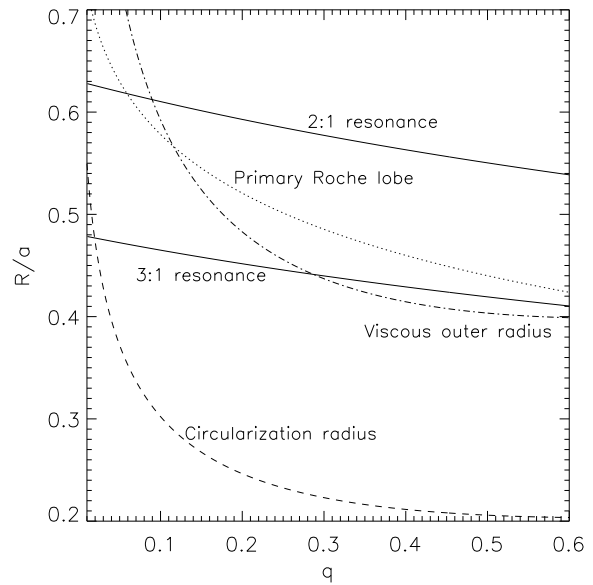}}}
\end{center}
\caption{Plot of resonant, Roche lobe, circularisation, and viscous disc radii about the primary star in a close binary as a function of mass ratio $q$.}
\label{fig:radii}
\end{figure}

\begin{figure*}
\begin{center}
\resizebox{88.1mm}{88.1mm}{
\mbox{
\includegraphics{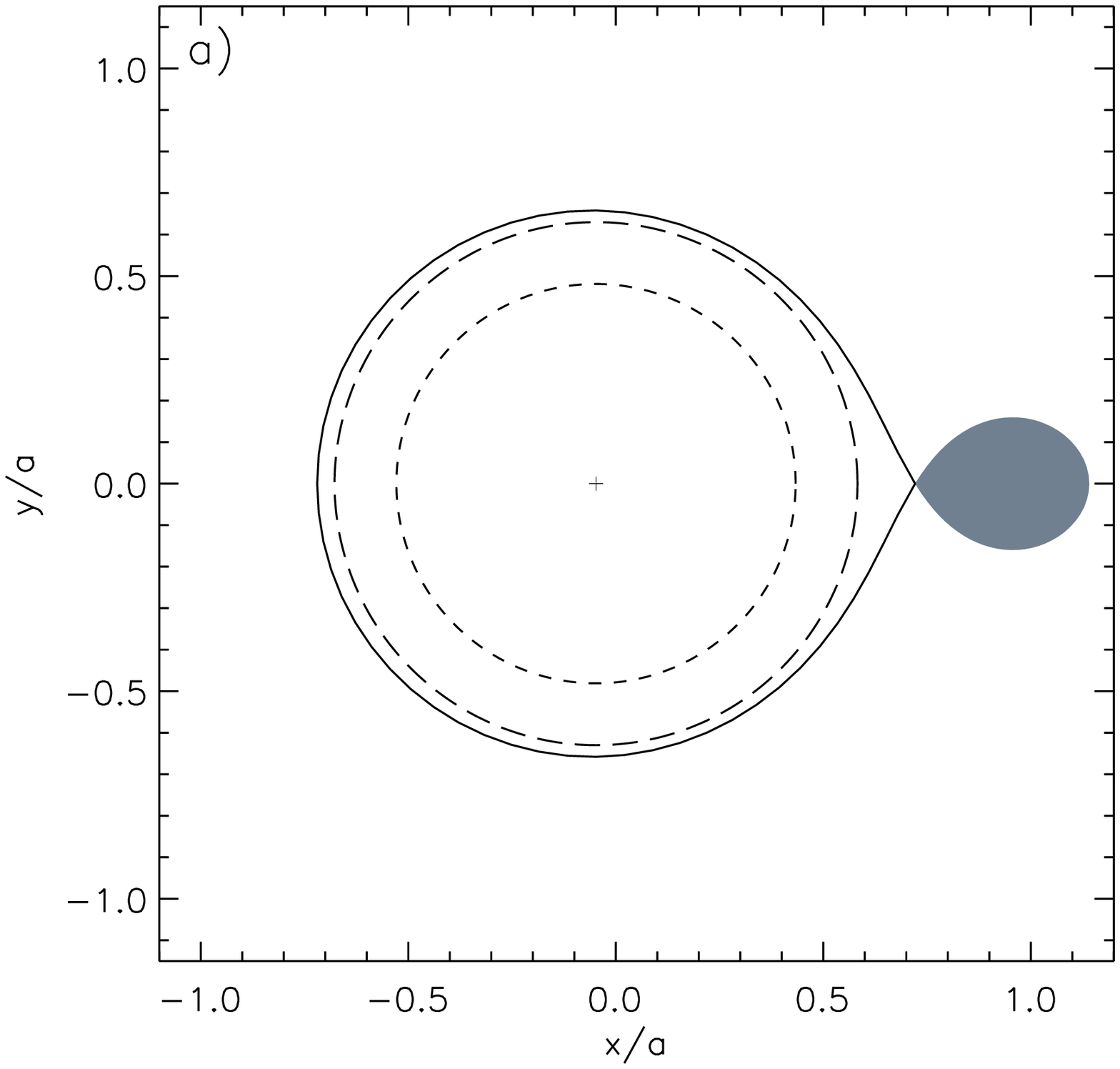}}} 
\resizebox{88.1mm}{88.1mm}{
\mbox{
\includegraphics{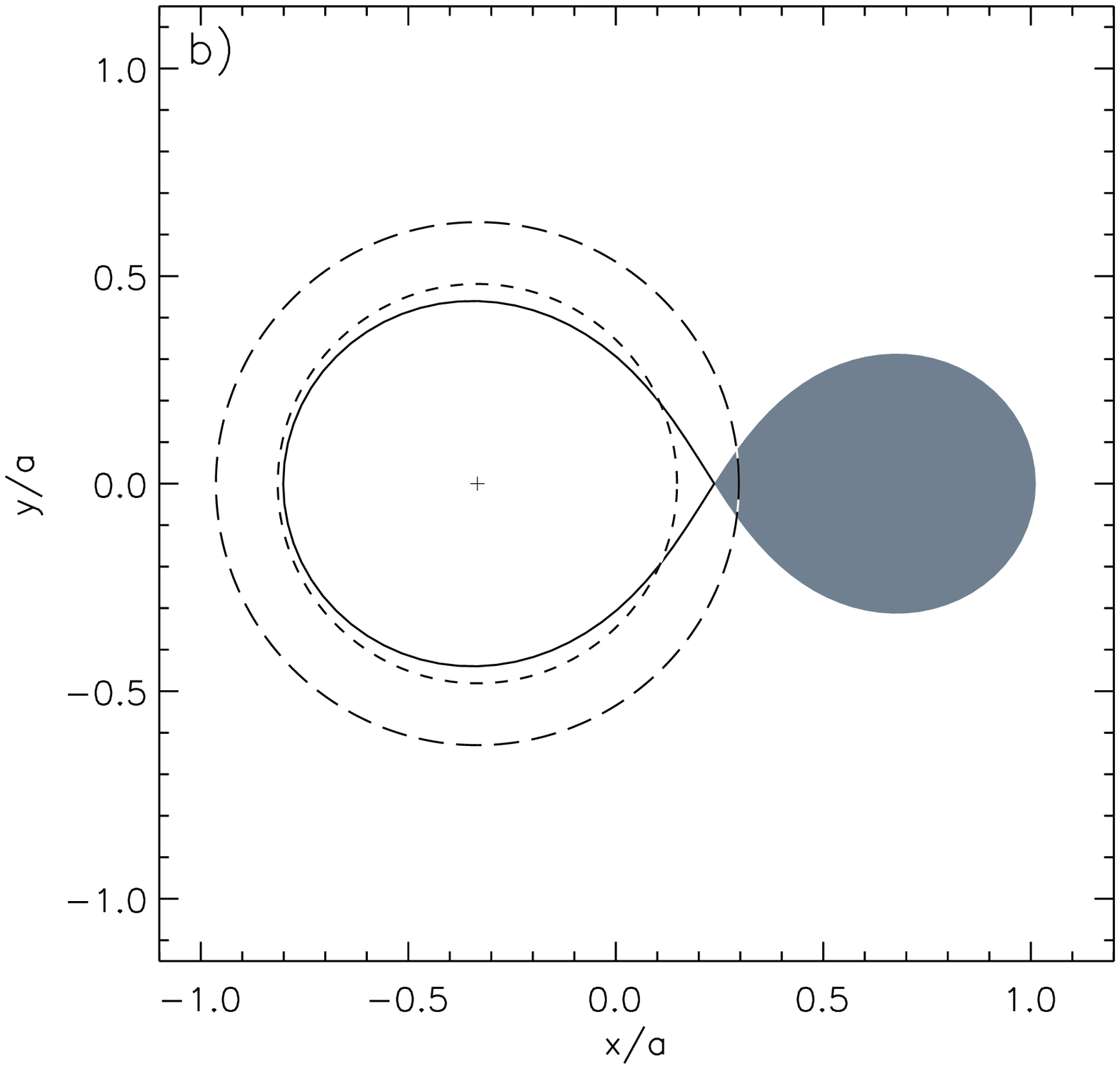}}}
\end{center}
\caption{This figure illustrates the positions of the 2:1 and 3:1 resonances in binaries with mass ratios of $q=0.05$ in panel a, and $q=0.5$ in panel b. The primary Roche lobe is marked with a solid line while the secondary star, which is assumed to be Roche lobe filling, is coloured grey. The two dashed circles represent the resonant radii, the inner circle representing the 3:1 resonance and the outer line being the 2:1 resonance. The length unit is the binary separation $a$ and the diagram is centred on the centre of mass. The primary star is represented by a cross. In case a, both the resonant radii are within the Roche lobe and accessible to the disc, whereas neither is available in case b.}  
\label{fig:res}
\end{figure*}

\section{Propeller-driven resonant mass loss}
\label{sec:bin}
\subsection{Orbital Resonances}
\label{sec:orb}
In this section we test the idea described in Section~\ref{sec:int} by establishing under which conditions the accretion disc has access to the orbital resonances. These conditions are conveniently described with reference to a series of important radii, measured from the primary star, all of which are plotted in Fig.~\ref{fig:radii}. Disc elements describe orbits that are approximately Keplerian, so that the positions of the orbital resonances are readily calculated from Kepler's law. Therefore, where $a$ is the binary separation, the 2:1 and 3:1 resonances are found respectively at
\begin{eqnarray}
\label{eqn:res21}
\frac{R_{2:1}}{a} & = & \left(\frac{1}{2}\right)^{2/3} \left( 1 + q \right)^{-1/3} \; ,\\
\label{eqn:res31}
\frac{R_{3:1}}{a} & = & \left(\frac{1}{3}\right)^{2/3} \left( 1 + q \right)^{-1/3} \; .
\end{eqnarray}
If an accretion disc is to have access to a resonance then the resonant radius must lie within the disc. If the disc radius can be related to the mass ratio then this consideration can be used to calculate $q_{\rm max}$, the maximum mass ratio for which resonant access is possible. There are several ways of estimating the extent of the disc. Firstly we use the Roche lobe radius. The Roche lobe of the accreting star provides a hard maximum for the disc radius. The disc is, in fact unlikely to extend this far in quiescence, but this provides a good estimate for the radius in outburst, when the disc is subject to increased viscous spreading. The Roche lobe radius is approximated by \citet{egg83} to be 
\begin{equation}
\label{eqn:rlobe}
%R_{\rm{L}} = 0.462 \left( \frac{1}{1+q} \right) ^{1/3} a ,\; \; {\rm{for}} \; q \geq 0.05 \; ,
\frac{R_{\rm{L}}}{a} = \frac{0.49 q^{-2/3}}{0.6 q^{-2/3} + {\rm{ln}} \left(1+q^{-1/3} \right)} \; ,
\end{equation}
where $q$ represents the mass ratio of the binary, $q = M_{2}/M_{1}$. Equation~\ref{eqn:rlobe} shows that the primary Roche lobe contracts with increasing $q$, as must the maximum extent of the accretion disc. Equating the 3:1 resonant radius and the Roch lobe radius, using equations (\ref{eqn:res21}) and (\ref{eqn:rlobe}), and solving for $q$ gives a maximum mass ratio for access to the 2:1 resonance thus
\begin{equation}
\label{eqn:q21}
q_{{\rm{max}},2:1}  =  0.06 \; , 
\end{equation}
The same technique when applied to the 3:1 resonance gives a maximum access ratio of $q_{{\rm{max}},3:1}  =  0.90$. \citet{pac77} computed $q_{{\rm max},3:1}$ by solving the restricted three-body problem, finding that
\begin{equation}
\label{eqn:q31b}
q_{{\rm{max}},3:1}  =  0.25 \; .
\end{equation}
This much smaller value for $q{{\rm{max}},3:1}$ reflects the fact that the accretion disc is tidally truncated within the primary Roche lobe. These results are also seen graphically in Fig. \ref{fig:radii} as the intersections of the resonant radii and the Roche lobe radius.  This approach therefore provides a good estimate of the maximum mass ratio $q_{\rm max}$ for which resonance access is permitted in outburst. For reference, the positions of the 3:1 and 2:1 resonances are also marked in Fig.~\ref{fig:res} for two typical CV mass ratios. 

We note that $q_{{\rm{max}},2:1}$ is precisely the mass ratio attributed to WZ Sge \citep{ski02}. Indeed an outburst orbital hump (OOH), probably driven by this resonance, was observed by \citet{pat02} before the onset of 3:1 superhumps during the 2001 outburst, suggesting that the disc can access this resonance during outburst, as would be expected. SU UMa stars exhibit superhumps due to the 3:1 resonance so one would expect them to obey condition (\ref{eqn:q31b}). Indeed, of all the confirmed SU UMa in \cite{rit98}, only two have $q > 0.25$. 

A rough estimate of the quiescent disc radius is given by the circularisation radius $R_{\rm cric}$. This is the radius of a circular orbit associated with a specific angular momentum identical to that with which mass crosses the first Lagrangian point. It is the radius at which a disc will begin to form, and so all discs must extend at least this far. The circularisation radius may be used as an estimate for the extent of a very cold disc and is given by \citep*[e.g.][]{fra01}
\begin{equation}
\label{eqn:rcirc}
\frac{R_{\rm circ}}{a} = \left(1+q\right) \left(\frac{R_{\rm L1}}{a}\right)^4
\; .
\end{equation}
Here, the position of $R_{\rm L1}$, the first Lagrangian point, may be found numerically, or it may be fitted by various expressions. The position of $R_{\rm L1}$ should not be confused with the $R_{\rm L}$, the {\it mean} Roche lobe radius as defined in equation (\ref{eqn:rlobe}). \citet{kop59} gives the following fit for $R_{\rm L1}$.
\begin{equation}
\label{eqn:L1}
\frac{R_{\rm L1}}{a} = 1 - \omega + \frac{1}{3} \omega^2 + \frac{1}{9} \omega^3
\; ,
\end{equation}
where
\begin{equation}
\omega^3 = \frac{q}{3 \left(1 + q\right)}
\; .
\end{equation}
Equating the 3:1 resonant radius and the circularisation radius using equations (\ref{eqn:res31}) and (\ref{eqn:rcirc}), and solving for $q$, gives $q_{\rm max,3:1} = 0.01$ as the maximum mass ratio for which the 3:1 resonance is accessible, if the disc is limited by the circularisation radius. Such a limit however implies that there is no outwards viscous spreading whatever. Even quiescent accretion discs are however subject to some viscous spreading. 

An improved estimate for the extent of the quiescent accretion disc is that proposed by \citet{osa02}. In this approach conservation of angular momentum in a steady-state viscous disc, {\it neglecting external torques},  is used to estimate the outer radius of an accretion disc thus
\begin{equation}
\label{eqn:r}
R_{\rm{out}} = \left( \frac{7}{5} \right) ^{2} R_{\rm circ} \; ,
\end{equation}
for a quiescent Shakura-Sunyaev disc. Combining this result with equation (\ref{eqn:res31}), this approach gives an estimate of the maximum mass ratio for 3:1 resonance access in the quiescent disc of
\begin{equation}
\label{eqn:q31}
q_{{\rm{max}},3:1} = 0.29 \; .
\end{equation}
We note however that access to the 2:1 resonance is still limited by the Roche lobe estimate given by equation \ref{eqn:q21} since the Roche lobe is smaller than the viscous limit in this regime, as shown in Fig.~\ref{fig:radii}.

\subsection{The magnetic propeller}
\label{sec:prop}
\begin{figure*}
\begin{center}
\resizebox{88.1mm}{88.1mm}{
\mbox{
\includegraphics{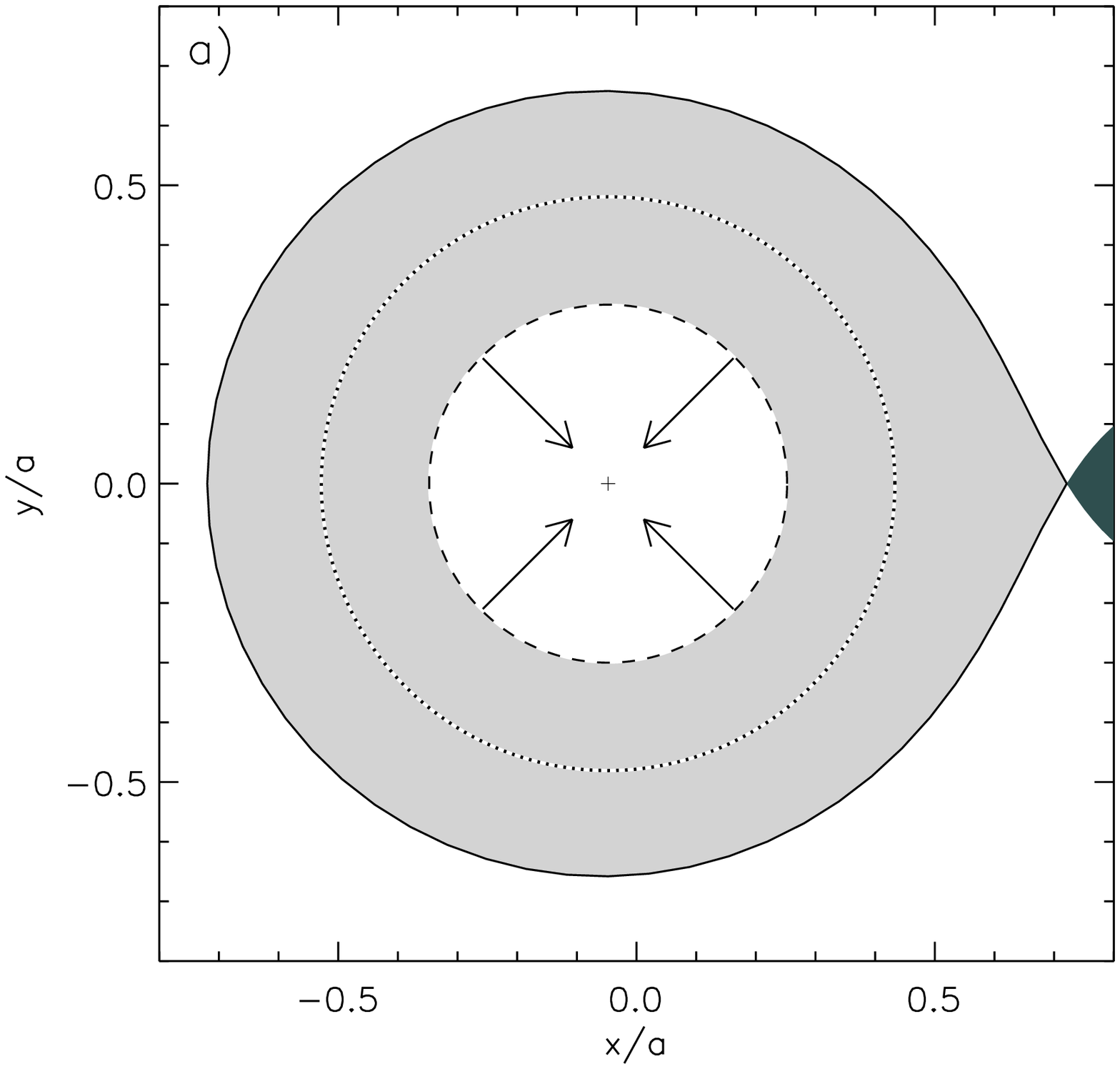}}} 
\resizebox{88.1mm}{88.1mm}{
\mbox{
\includegraphics{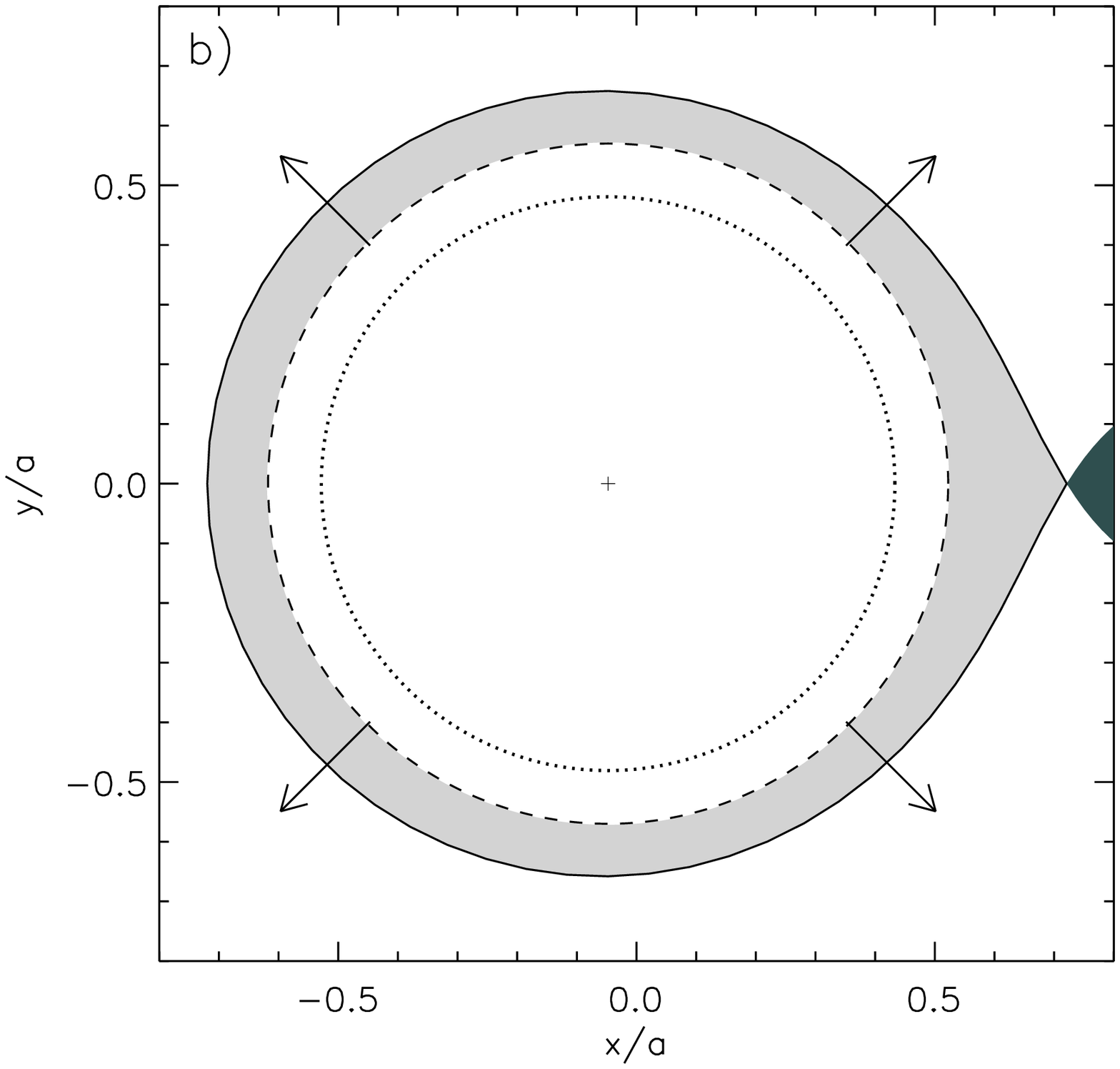}}}
\end{center}
\caption{This figure illustrates magnetically enhanced accretion in a, and the magnetic propeller in b. The primary Roche lobe is marked with a solid line while the secondary star, which is assumed to be Roche lobe filling, is coloured dark grey. The accretion disc is shaded light grey and is bounded at its inner edge by a dashed circle representing the magnetic radius. The dotted line denotes the co-rotation radius. The length unit is the binary separation $a$ and the diagram is centred on the centre of mass. The primary star is represented by a cross. The difference between the two cases is explained fully in the text.}  
\label{fig:rco}
\end{figure*}

So far we have discussed access to orbital resonances for an unperturbed, quiescent disc. During outburst the disc will expand outwards due to increased viscous angular momentum transport, hence the detection of superhumps in superoutbursts. Another way to aid access to the binary resonances is by the addition of an additional source of angular momentum, such as a magnetic torque. A magnetic field on the primary star can act either to drive disc matter inwards {\it or} outwards; it may behave either as a source or as a sink of angular momentum. It is useful here to examine a simplified case where the primary magnetic field rotates as a rigid body, rooted to the primary star and rotating with angular frequency $\Omega_{\star}$. The magnetic or magnetospheric radius $R_{\rm{mag}}$ is the radius at which the magnetic time-scale is equal to the viscous time-scale; in other words it is the largest radius at which magnetic torques are significant in the disc. The co-rotation radius $R_{\rm{co}}$ is the radius at which an element of disc, in describing a Keplerian orbit about the primary star, shares the angular frequency of the star and the magnetic field. At this co-rotation radius, analogous to that of a geostationary orbit, there is no drag force between the field and the disc, since they have the same velocity. However $R_{\rm{co}}$ marks the boundary between the magnetic accretion and magnetic propeller regimes, as illustrated in Fig.~\ref{fig:rco}.

\begin{figure}
\begin{center}
\resizebox{80.0mm}{80.0mm}{
\mbox{
\includegraphics{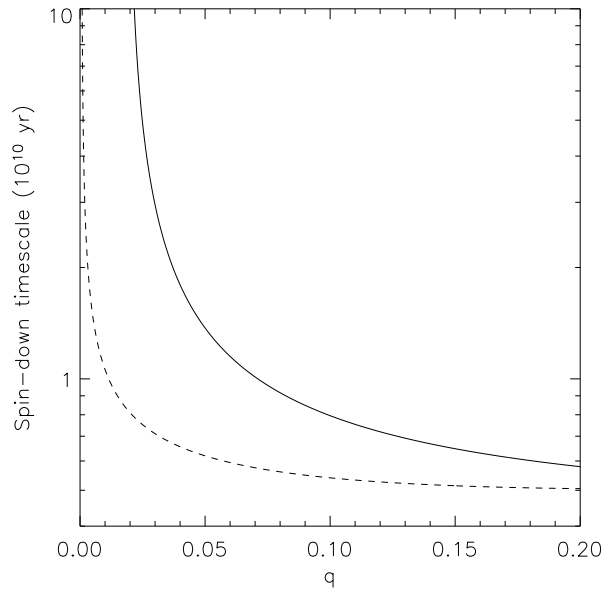}}}
\end{center}
\caption{Spin-down time-scale $\tau$ as a function of mass ratio $q$ for a typical CV. In the resonant case, plotted with a solid line, the time-scale increases asymptotically as the circularisation radius approaches the 3:1 resonance while $\tau_{\rm spin} \sim 10^{10} \ {\rm yr}$ for most $q$. In the non-resonant, or strong propeller case, plotted with a dashed line, the spin-down time-scale becomes large only for $q < 0.005$.}
\label{fig:graham}
\end{figure}

At those radii where $R < R_{\rm{co}}$ the magnetic field rotates less rapidly than the orbiting disc, and so acts to brake it, aiding accretion. If $R_{\rm{mag}} < R_{\rm{co}}$ therefore, then the disc will be be disrupted at $R_{\rm{mag}}$ by the effect of rapid magnetic accretion, causing the star to spin up, as is the case in an intermediate polar. However, if $R_{\rm{mag}} > R_{\rm{co}}$ then then disc material, spreading inwards, first encounters a significant magnetic torque where the field is travelling {\it{faster}} than the disc, causing the transfer of angular momentum from the star to the disc, pushing the disc outwards and spinning the star down. Although magnetic accretion and the magnetic propeller both result in a `hole' in the accretion disc, they are distinct mechanisms with different outcomes for the primary spin and for the disc structure. Disc truncation by the magnetic propeller is implicated in the long recurrence time of WZ~Sge, and in FU~Ori young stars \citep*{mat03}. AE~Aqr is an extreme case of propeller truncation.

In outbursting CVs, such as dwarf novae, an outburst cycle exists and a steady state is never reached. A propeller may reduce the disc mass sufficiently so that such an outburst is never triggered and a steady state is reached. In this steady state, the disc must lose mass at the same rate at which it is fed by the accretion stream. This can lead to only two possible steady-state solutions. The disc will spread viscously as mass is added and either the inner disc will reach $R_{\rm{co}}$ and the system will become a magnetic accretor, or the outer disc will extend beyond $R_{\rm{L}}$ and mass will be lost from the Roche lobe. In the latter case, that of the true propeller, a strong magnetic field is required, since mass must be pushed to radii beyond $R_{\rm{L}}$, and a rapid spin down is therefore suffered by the primary star. Such a state cannot therefore be long lived and such systems are therefore likely to be rare, and will not be significant in the CV period distribution.

\subsection{Combined effect of resonances and propeller}
The model proposed in this paper uses both magnetic and resonant torques to enable a weak magnetic propeller to eject mass from the primary Roche lobe. The model requires the magnetic propeller to push mass only as far as the 3:1 resonance. This implies a much weaker propeller, and hence a longer spin down time. Such a propeller state may therefore exist for much longer than one which relies on the magnetic torque alone. In this picture, disc elements that are propelled to the resonant radius will be excited into eccentric orbits, losing angular momentum. The newly eccentric orbits may carry disc material outside the Roche lobe at apastron, where it will be ejected from the binary or re-accreted by the secondary star. Both the propeller and the resonant torque are critical to this model.

Propeller-driven resonant mass loss will occur if two conditions are satisfied. Firstly, the circularisation radius must also lie close to the 3:1 resonance so that a weak magnetic torque can lift mass to the resonant radius. Secondly, the 3:1 resonance should lie within, but ideally close to, the Roche lobe radius so that a small eccentricity is able to drive mass from the Roche lobe and cause mass loss. 

The magnetic torque and hence the white dwarf spin-down time-scale required to drive resonant access can be estimated by calculating the torque required to lift mass from the circularisation radius to the 3:1 resonant radius, assuming that such mass will then be lost to the Roche lobe as a result of resonant torques. In the steady state the rate at which mass is transferred from $R_{\rm circ}$ to $R_{\rm L1}$ is identical to the binary mass transfer rate $\dot{M}$. This gives a magnetic torque, which in the absence of accretion or any other sources of angular momentum, is identical to the primary spin-down torque, of
\begin{equation}
\label{eqn:magtorque}
\dot{J}_{\rm mag} = \sqrt{G M_{1}} \dot{M} \left(R^{1/2}_{3:1} - R^{1/2}_{\rm circ} \right) \; ,
\end{equation}
where $G$ represents the constant of universal gravitation and $M_{1}$ is the mass of the primary star. Equations (\ref{eqn:rcirc}) and (\ref{eqn:magtorque}) may be combined to give
\begin{equation}
\dot{J}_{\rm mag} = \sqrt{G M_{1}} \dot{M} a  \left( \gamma^{1/2} - \left[1 + q \right]^{1/2} \left[\frac{R_{\rm L1}}{a} \right]^{2} \right) \; ,
\end{equation}
where $R_{\rm L1}$ is obtained for example from equation (\ref{eqn:L1}) and where $\gamma = 0.48$ for the 3:1 resonance. The spin-down time-scale is given by
\begin{equation}
\label{eqn:tspin}
\tau_{\rm spin} = \frac{J_{\rm wd}}{\dot{J}_{\rm mag}} = \frac{2 \pi I_{\rm wd}}{P_{\rm spin} \dot{J}_{\rm mag}}
\end{equation}
where $J_{\rm wd}$ is the spin angular momentum of the white dwarf, $I_{\rm wd} \sim M_{1} R_{\rm wd}^{2}$ is the moment of inertia, $R_{\rm wd}$ is the radius of the white dwarf and $P_{\rm spin}$ is the WD spin period. The spin-down time-scale is then takes the form
\begin{equation}
\tau_{\rm spin} = \frac{2 \pi R^2_{\rm wd} M_{1}^{1/2} }{P_{\rm spin} \dot{M} G^{1/2} \left(R^{1/2}_{3:1} - R^{1/2}_{\rm circ} \right)} \; .
\end{equation}
The spin-down time-scale is then a weak function of $a$ and $M_{1}$, but a strong function of $q$, rising asymptotically at $R_{3:1} = R_{\rm circ}$. For a CV with orbital period $P_{\rm orb} = 80 \ {\rm min}$, primary spin period $P_{\rm spin} = 30 \ {\rm s}$, $R_{\rm wd} = 1 \times 10^{9} \ {\rm cm}$, mass transfer rate $\mdot = 3 \times 10^{-11} \ \msun \ {\rm yr}^{-1}$ and primary mass $M_{1} = 1.2 \msun$ (selected to mtach numerical simulations) the spin-down time-scale is roughly constant at $\tau_{\rm spin} \sim 10^{10} \ {\rm yr}$ for most $q$, but is much longer for $q \lesssim 0.05$. This is illustrated in Fig.~\ref{fig:graham} with an asymptote at $q \sim 0.02$. For $q \lesssim 0.02$ the circularisation radius lies outside the 3:1 resonant radius so that only the inner disc would have access to the torque from this orbital resonance. This may explain the results of \citet{sim98} who find, in SPH simulations of non-magnetic systems, that an eccentric precessing disc does not form where $q \lesssim 0.025$. \citet{sim98} themselves invoke the damping effect of the 2:1 resonance to explain this effect.

For comparison, in the case of a pure magnetic propeller, where resonant torques are not considered, mass must be lifted by the magnetic torque all the way from the circularisation radius to the Roche lobe surface. Applying the same method as above this gives
\begin{equation}
\label{eqn:magtorquenores}
\dot{J}_{\rm mag} = \sqrt{G M_{1}} \dot{M} \left(R^{1/2}_{\rm L} - R^{1/2}_{\rm circ} \right) \; .
\end{equation}
The spin down time can be computed in the same way, and exhibits the same asymptotic behaviour. However, as shown in Fig.~\ref{fig:graham}, the spin-down time exceeds $1 \times 10^{10} \ {\rm yr}$ for the non-resonant case only for unphysical mass ratios where $q < 0.005$.

In the standard picture of CV evolution, a large number of systems are expected to accumulate around the period minimum, with small mass ratios. The mechanism outlined above effectively halts accretion onto the primary star for a significant fraction of the CV lifetime. During this phase outbursts are prevented and the disc mass is greatly reduced, also reducing the disc luminosity and making the system effectively undetectable. This phase will be much longer lived for low $q$ CVs so that more short period CVs would be hidden from view, providing a possible explanation for the absence of the period spike.

\section{Smoothed Particle Hydrodynamics Calculations}
\label{sec:num}
The smoothed particle hydrodynamics (SPH) method is a Lagrangian particle technique for the modelling of fluid dynamics. The fundamental principle of SPH is that fluid properties such as pressure and density are carried by particles which represent elements of the fluid in question. The values of these properties may then, at any point in space, be interpolated using a weighted mean of the values carried by nearby particles. Conservation of angular momentum is inherent in the SPH technique, which makes it an ideal method for the modelling of accretion discs. A review of the SPH method in general can be found in \citet{mon92} while the particular code used here is discussed by \citet{mur96} and by \citet{tru00}. The viscosity used in this work is the linear element of the SPH viscosity described by \citet{mon92}. In the divergence free case, this viscosity can be shown to be equivalent to the Navier-Stokes shear viscosity and to the \citet{sha73} alpha disc viscosity \citep[\eg][]{pon88,mat04th}. In three dimensions this viscosity is given by
\begin{equation}
\label{eqn:visc}
\nu = \frac{1}{10} \zeta c_{\rm{s}} H \sim \alpha c_{\rm{s}} H \; ,
\end{equation}
where $\zeta$ is the SPH linear viscosity parameter, $\alpha$ is the Shakura-Sunyaev viscosity parameter, $c_{\rm{s}}$ is the sound speed in the disc and $H$ is the disc scale height. In this version of the code the sound speed is fixed throughout the calculation. 

Full magnetohydrodynamics have not yet been successfully incorporated into SPH though efforts have been made in this direction \citep[e.g.][]{pri05}. A simple approach to modelling the effect of the primary magnetic field is to apply a magnetic torque prescription to the SPH particles. The magnetic element of the momentum equation can be expressed as
\begin{equation}
\label{howsya}
{\bf{J}} \times {\bf{B}} = \frac{1}{\mu_{\rm r}}  \left( {\bf{B}} \cdot \nabla \right) {\bf{B}} - \nabla \left( \frac{{\bf{B}}^{2}}{2 \mu_{\rm r}} \right)
\; ,
\end{equation}
where ${\bf{J}}$ is current density, ${\bf{B}}$ is magnetic field and $\mu_{\rm r}$ is the permeability of the plasma. The first term on the right hand side is in the form of a tension, which vanishes if the field lines are not curved, and the second represents a pressure which can usually be neglected for slowly varying fields. 

If equation (\ref{howsya}) is transformed into cylindrical polar coordinates, and if it is assumed that the radial component of the magnetic field is small and that $B^{2} \sim B_{\rm{z}} B_{\rm{\varphi}}$, then the magnetic acceleration takes the form \citep[\eg][]{mat04th}
\begin{equation}\label{nneweq2}
\amag \sim \frac{1}{\rho \rc} \left( \frac{\bz \bphi}{4 \pie} \right)
\; ,
\end{equation} 
where $\rho$ is density and $r_{\rm{c}}$ represents the radius of curvature of the magnetic field lines, while $B_{\rm{z}}$ and $B_{\rm{\varphi}}$ represent the vertical and azimuthal components of the magnetic field respectively. We make the further approximation that the radius of curvature of the field lines is of the order of the disc scale-height so that $\rc \sim H$
\citep*[e.g.][]{pea97}. The ratio of vertical to azimuthal field strengths is
related to the shear between the disc and the magnetic field. If it is
assumed that the field rotates rigidly with the star then this ratio can be
expressed in the form \citep[\eg][]{liv92} 
\begin{equation}\label{nneweq3}
\frac{\bphi}{\bz} \sim - \frac{\left(\omegak - \omegastar \right)}{\omegak}
\; ,
\end{equation}
where $\Omega_{\rm{k}}$ denotes the Keplerian angular velocity at radius $R$ from the star. If Keplerian orbits are assumed, the surface density related to the disc scale height by $\Sigma \sim \rho H$, and the magnetic moment is defined by the expression $\mu = B_{\rm z} R^{3}$, then we may make the approximation
\begin{equation}\label{nneweq5}
a_{\rm{mag}} \sim -\frac{{\bf{\mu}}^{2} R^{-6}}{4 \pi \Sigma} \frac{\left(\omegak - \omegastar \right)}{\omegak} \sim - \frac{{\bf{{\mu}}}^{2} R^{-6}}{4 \pi \Sigma} \frac{\left|{\bf{v}}_{\rm{d}} - {\bf{v}}_{{B}} \right|}{{v}_{\rm{d}}}
\; , 
\end{equation}
where ${\bf{v}}_{\rm{d}}$ and ${\bf{v}}_{B}$ are the velocities of the disc and the magnetic field respectively. It is an acceleration of this form which is added to the particles in the SPH model to represent the effect of the primary magnetic field. The acceleration is fixed in the code by the parameter 
\begin{equation}
\label{eqn:km}
k_{\rm{m}} = \frac{{\bf{\mu}}^{2}}{4 \pi \Sigma} \; .
\end{equation}
The effects of the binary resonance require no such effort to reproduce, but emerge naturally as a result of gravitational forces exerted on the particles by the binary. Every particle is subject to a gravitational acceleration from each star in addition to the magnetic acceleration defined in equation (\ref{nneweq5}) and fluid dynamic forces. In order to compute the predicted spin evolution of the primary star, the torques due to magnetic and gravitational accelerations are calculated and recorded separately. However, at this stage the code permits neither the spin of the primary nor the orbital period to evolve as a result of the computed angular momentum transfer.

Mass is injected at the first Lagrangian point and is removed from the simulation in three ways. Particles which reach the surface of the white dwarf at $R_{\star} = 6 \times 10^{8} \ {\rm{cm}}$ are removed and count towards accreted mass. If elements of the disc extend beyond $1 \times 10^{10} \ {\rm{cm}}$ then they removed and counted as mass lost to the system. If mass re-enters the secondary Roche lobe, then it is considered to have been re-accreted by the secondary star.

\begin{figure*}
\begin{center}
\resizebox{88.1mm}{88.1mm}{
\mbox{
\includegraphics{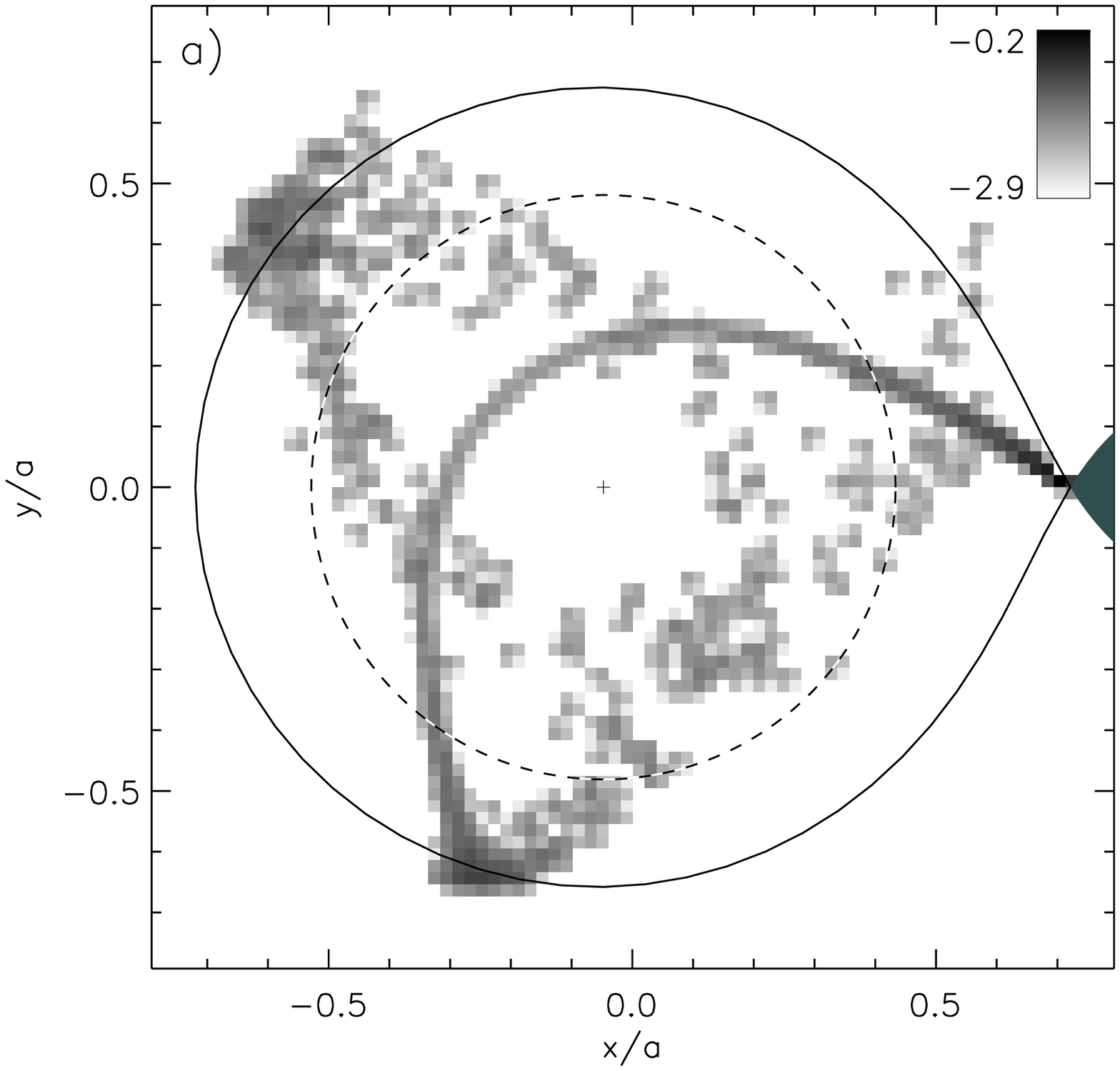}}} 
\resizebox{88.1mm}{88.1mm}{
\mbox{
\includegraphics{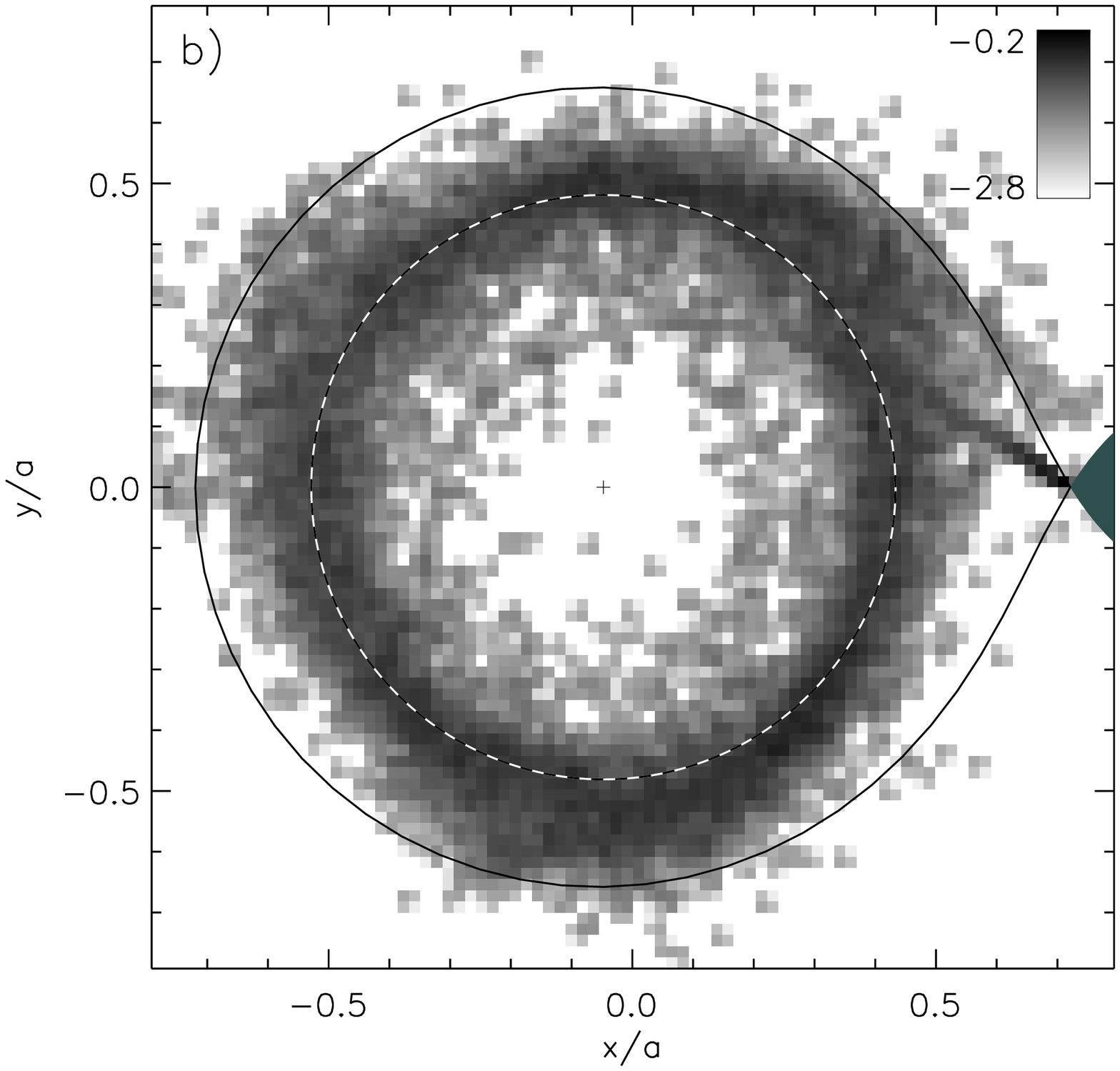}}} 
\resizebox{88.1mm}{88.1mm}{
\mbox{
\includegraphics{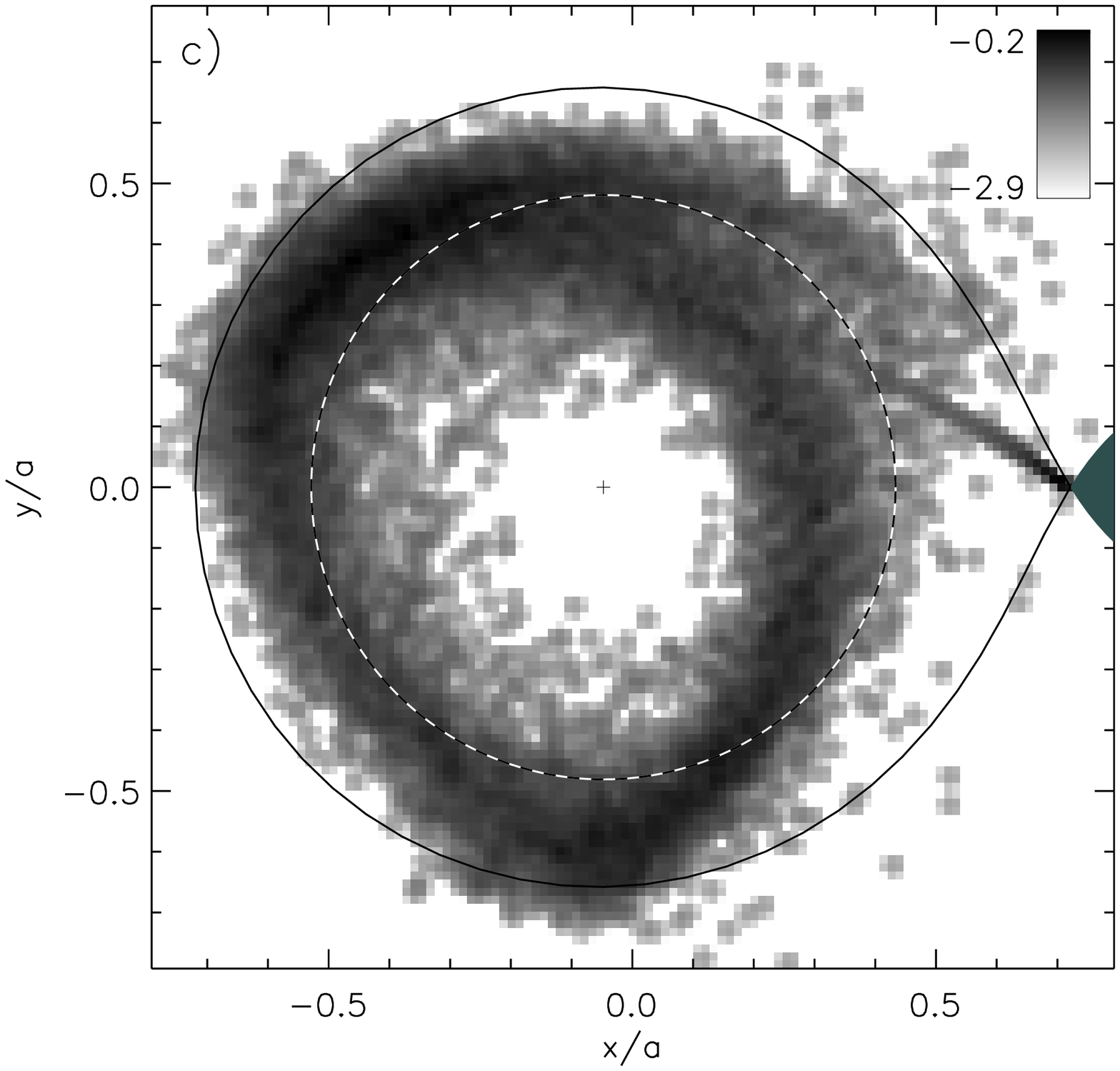}}}
\resizebox{88.1mm}{88.1mm}{
\mbox{
\includegraphics{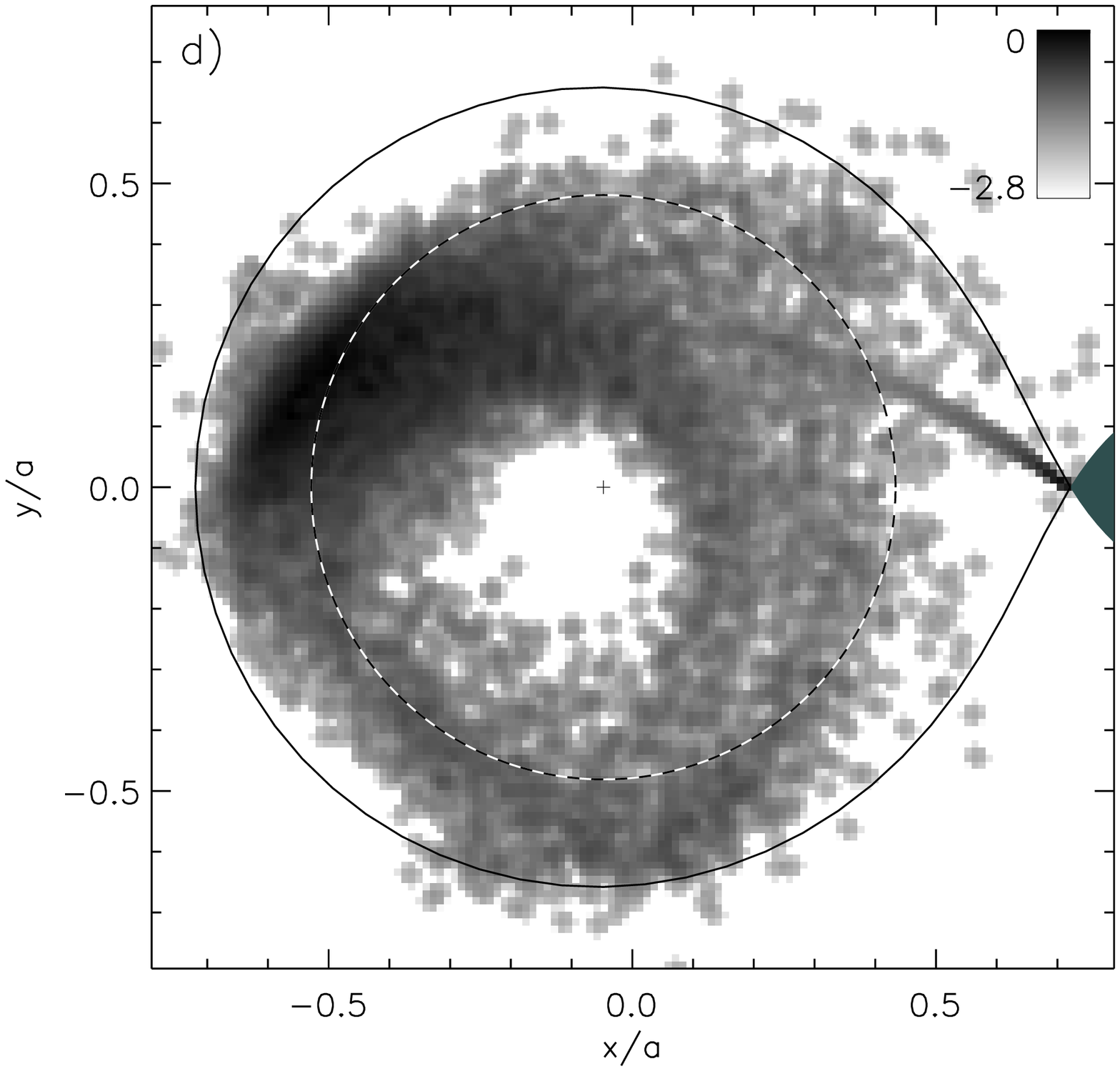}}}
\end{center}
\caption{These plots, taken from an SPH computation, of log surface density in ${\rm{g}} \ {\rm{cm}}^{-1}$ show how propeller-driven resonant mass ejection might begin, starting from an empty disc. The simulation contained $\sim 30,000$ particles with a sound speed of $c_{\rm s} \sim 1 \times 10^{6} \ {\rm cm \ s^{-1}}$. The primary Roche lobe is marked with a solid line while the secondary star, which is assumed to be Roche lobe filling, is coloured grey. The dashed circle represents the 3:1 resonant radius. The length unit is the binary separation $a$ and the diagram is centred on the centre of mass. The binary is orbiting in the anticlockwise (direct) sense. The primary star is represented by a cross. This simulation was performed with a mass ratio of $q=0.05$, a primary mass of $M_{1}=1.2 \ \msun$, a binary period of $P_{\rm{orb}}=80 \ {\rm{min}}$, a primary spin period of $P_{\rm{spin}}=30 \ {\rm{s}}$, a mass transfer rate in the accretion stream of $\mdot=3.0 \times 10^{-11} \ {\rm{\msun \ yr^{-1}}}$ and a surface magnetic field of $5 \ \rm{kG}$. In frame a, after a single orbit, the disc is just beginning to fill and the initial rosette structure is still visible. In frame b, after $15$ orbits, the disc has become circularised as a result of viscous processes. However, in frame c, after $25$ orbits, the resonance is beginning to drive an eccentricity, and by frame d, after only $50$ orbits, that eccentricity is sufficient to cause extensive mass loss, with a large amount of mass being swept up by the secondary star.}
\label{fig:trunc}
\end{figure*}

\section{Numerical results}
\label{sec:res}
A series of numerical experiments were performed to confirm the viability of  the mechanism explained in Section~\ref{sec:bin} and to briefly explore the parameter space for which that mechanism is operable. Certain basic parameters of the system were fixed: the total mass of the system was set to $M_{\rm{tot}} = 1.2 \ \msun$, because the simulations were initially part of a programme to model WZ~Sge. The orbital period was $P_{\rm{orb}} = 80 \ {\rm{min}}$ and the spin period of the primary star was a deliberately short $P_{\rm{spin}} = 30 \ {\rm{s}}$. Particles were injected into the primary Roche Lobe at a rate of $\mdot = 3.0 \times 10^{-11} \ {\rm{\msun \ yr^{-1}}}$ although in some cases little of this mass found its way to the primary star. The Shakura-Sunyaev alpha viscosity was fixed at $\alpha = 0.01$ throughout the calculations. In all cases the computation began with an empty disc and was permitted to run until a steady state was reached. In reality such systems would evolve, with full discs, from longer periods but the choice of an empty disc as an initial condition provides a clear illustration of the onset of resonant mass ejection and, since such simulations can be performed rapidly, allows a greater number of computations to be performed. The particle mass was varied between models to enable steady state to be achieved in those computations which led to high mass discs, without causing lower mass discs to be under resolved. 
%At steady state each disc contained $N \gtrsim 2 \times 10^{4}$ particles and convergence tests showed the results presented here to be robust to changes in resolution. This is unsurprising since the effect seen is predominantly a dynamical rather than a fluid one.

\begin{figure}
\begin{center}
\resizebox{88.1mm}{88.1mm}{
\mbox{
\includegraphics{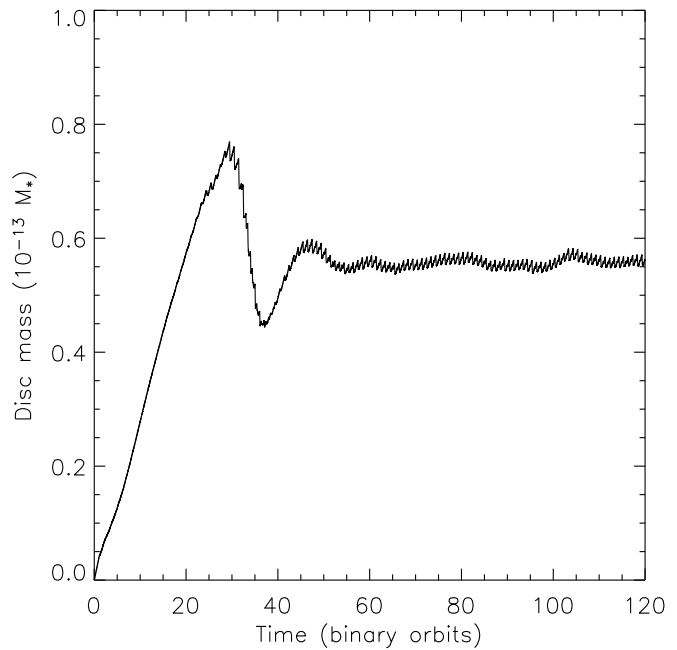}}}
\end{center}
\caption{This figure shows the variation of disc mass with time during the onset of propeller-driven resonant mass ejection, beginning from an empty disc. Results are taken from the same SPH computation as for Fig.~\ref{fig:trunc}. Extensive mass loss commences after 40 binary orbits and a quasi-steady state is then reached. From this time onwards a modulation appears, at a frequency close to that of the binary orbit. Note the agreement between this result and that shown in fig. 5 of \citet{hir90}.}  
\label{fig:graphs}
\end{figure}

\begin{figure}
\begin{center}
\resizebox{88.1mm}{88.1mm}{
\mbox{
\includegraphics{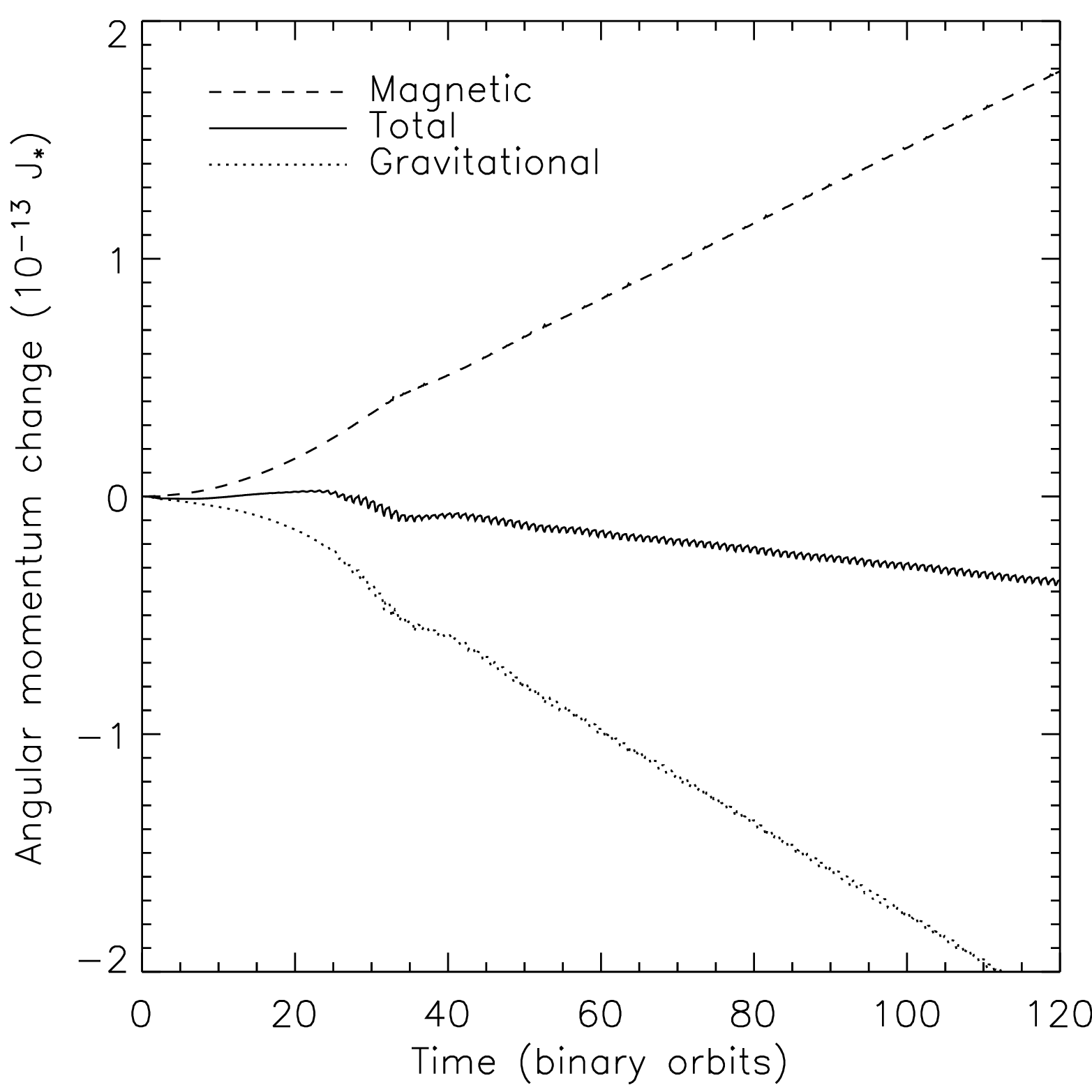}}}
\end{center}
\caption{This figure shows the change of angular momentum due to magnetic and gravitational torques during the onset of propeller-driven resonant mass ejection, beginning from an empty disc. Results are taken from the same computation as for Figs. \ref{fig:trunc} and \ref{fig:graphs}. The effect on disc angular momentum due to mass flux is not illustrated.}  
\label{fig:graphs2}
\end{figure}

The experiment was performed with mass ratios of 0.05, 0.1, 0.2, 0.4 and 0.8 and the magnetic parameter $k_{\rm{m}}$ was set to values which, for a primary star of radius $R_{\star} \sim 1 \times 10^{9} \ {\rm{cm}}$ and a disc surface density of $\Sigma \sim 1 \ \rm{g \ cm^{-2}}$, correspond to very modest magnetic field strengths of $B_{\rm z} = 5 \ {\rm{kG}}$, $4 \ {\rm{kG}}$, $2 \ {\rm{kG}}$ and zero at the surface of the white dwarf. It should be noted however that the relationship between the magnetic field strength and the code parameter $k_{\rm{m}}$ is model dependent, so that results should be treated as no more than a guide to the real fields required to support this mechanism. 

The results showed that both the steady-state disc mass and primary accretion rate were an order of magnitude lower in the magnetic cases than in the non-magnetic case; the disc mass fell from $~10^{-13}$ to $~10^{-13}$ of the binary mass while the accretion rate onto the primary fell from $~0.9$ to $~0.1$ of the mass transfer across the L1 point. Neither the magnetic field strength amongst the magnetic cases, nor the mass ratio had a clear effect on the steady-state disc mass or on the white dwarf accretion rate. However, these simulations were performed at mass ratios higher than those for which a strong dependence would be expected.

%The results of these calculations are shown in Fig.~\ref{fig:table}, which shows how the WD accretion rates and disc mass vary with the mass ratio and the magnetic field. The first panel shows the mass of the accretion disc as a function of $q$ and the second panel shows the proportion of mass crossing the first Lagrangian point which is accreted by the primary. The rest of the material is either lost to the system or, as is the case for the vast majority of the mass, is re-accreted by the secondary. Propeller-driven resonant mass loss is most pronounced at low mass ratios, as expected, and is substantial only in the case where a magnetic field is present. This is because in systems with low mass ratios, smaller torques are required to access the 3:1 resonance, as discussed in Section~\ref{sec:bin}. Increasing the magnetic field strength within the simulated (non-zero) range does not greatly influence mass ejection rates at low mass ratios, where resonant access is possible with a weak field. However the magnetic field strength has a more significant effect at higher mass ratios, where the resonant radius is more difficult to reach, so that a stronger field and hence a {\it more rapid spin-down} is required to prevent accretion.

An example of resonant mass ejection, drawn from the range of calculations described above, is illustrated in Fig.~\ref{fig:trunc}. Disc matter is prevented from accreting by the magnetic propeller and the disc spreads outwards. When the disc grows sufficiently to be able to access the 3:1 resonance, eccentric modes are excited in the disc. The eccentricity grows until a proportion of the disc protrudes beyond the primary Roche lobe and can be swept up by the secondary star, or is lost to the system. That this eccentricity is not the result of initial particle velocities is clear from the fact that the disc passes through a circular phase and that the effect is absent in the non-magnetic case.

The longer-term evolution of the disc in the same computation is illustrated in Figs.~\ref{fig:graphs} and \ref{fig:graphs2} where the disc mass and the torque exerted upon the disc are plotted against time for around a hundred orbits. It is clear that once the disc `catches' the 3:1 resonance after about 40 orbits, disc mass is lost rapidly from the system until a quasi-steady state is reached. This result is in good agreement with simulations performed by \citet{hir90}. Both the disc mass and the gravitational torque upon the disc exhibit a modulation on roughly the orbital period, which emerges at the same time as the aforementioned dramatic mass loss episode. This is easily understood by considering the secondary star sweeping up disc mass once per orbit from the protruding part of the disc, and exerting a similarly modulated resonant torque. In fact, taking the precession of the disc into account, the modulation of the disc mass would be expected to occur on the beat frequency between the binary orbit and the disc precession period: the superhump period. This indeed proves to be the case. No such modulation is observed in the magnetic torque which remains roughly constant with time after the quasi-steady state has been reached. This is as expected, because the magnetic field acts predominantly on the inner edge of the accretion disc.

By definition, the total angular momentum of the disc does not exhibit long term variation after the quasi-steady state has been reached. The gravitational torque acts to counteract that applied by the primary magnetic field. The disc is spun up by the magnetic field and spun down again by the binary, so that a weak spin couple exists between the binary orbit and the primary spin. This drives a long term tendency towards a synchronous lock, but on a time-scale so long as to be largely irrelevant. Fig.~\ref{fig:graphs2} shows a small net torque acting to spin down the disc. However, the plot shows only applied torques and does not include the flux of angular momentum carried by particles entering and leaving the computation, which counteracts this trend, and allows a quasi-steady state to be maintained.

An approximate spin-down time-scale for the primary star can be calculated from equation (\ref{eqn:tspin}) where $\dot{J}_{\rm{mag}}$ is calculated here from numerical results, disregarding accretion torques. A value may be taken from Fig.~\ref{fig:graphs2} for example. In all cases where accretion was prevented in the above computations, spin-down times are found to be of the order $t_{\rm{spin}} \sim 10^{10} \ {\rm{yr}}$ in good agreement with analytic treatment in the previous section for $q \lesssim 0.05$. Numerical simulations have not yet been performed for very small $q$ values, but analytic results show that the spin-down time-scale would be much longer.

\section{Evolutionary consequences}
\label{sec:dis}
It has been shown that accretion, and hence luminosity, may be greatly reduced in CVs by a weak propeller acting together with orbital resonances. In this mechanism the magnetic field of the rapidly spinning primary drives matter to the 3:1 orbital resonance. The action of this resonance then excites eccentric modes in the disc material. Disc mass is then lost from the primary Roche lobe and some may be re-accreted by the donor star. A CV may exist in this state for a significant fraction of its lifetime. Importantly, this non-accreting state can last much longer in systems close to the period minimum, so that a larger proportion of short period CVs would be expected to exist in this invisible state. This mechanism of propeller-driven resonant mass ejection may therefore go some way towards explaining the absence of the predicted excess of short period CVs; the period spike.

In this picture, CVs evolving towards shorter periods will enter the non-accreting weak propeller state at various mass ratios, depending on the spin and magnetic field of the accreting white dwarf. This state is reached as soon as the 3:1 orbital radius becomes accessible to magnetically torqued disc matter. This phase will continue until the white dwarf has spun down sufficiently that accretion can resume, which occurs on a time-scale of $\tau_{\rm spin} \sim 10^{10} \ {\rm yr}$ for most mass ratios but which increases greatly at very small $q$. Because the non-accreting state is longer lived in short period systems we would expect this mechanism for hiding the period spike to be more efficient for those CVs entering this weak-propeller state later in their binary evolution, i.e. those with weaker primary magnetic fields. If propeller-driven resonant mass ejection does occur in CVs then it will greatly reduce the luminosity due to the accretion disc and to accretion onto the white dwarf. A population of low-luminosity CV-like objects with low luminosity is therefore predicted by this model, for which there are currently no known candidates. However, since in addition to a greatly reduced luminosity, outbursts would also be prevented in such systems, the non-detection of these objects may be due to selection effects. If such stars were detected however, then they would be expected to exhibit permanent superhumps, since the disc would always be eccentric and precessing. It may also be possible to observe the re-accretion of matter by the secondary star in these objects.

In normal CV evolution, mass transfer from the secondary to the primary star, driven by an angular momentum loss mechanism such as gravitational radiation, reduces the mass ratio with time. In the weak propeller mechanism proposed here, most mass is not accreted by the primary star, but instead is re-accreted by the secondary so that the mass ratio does not evolve. For the secondary Roche lobe to accept this mass, either the contraction of the Roche lobe must be slowed or the secondary star itself must be made to contract more rapidly. The weak propeller mechanism lends itself to both of these possibilities. Matter is returned to the secondary with increased angular momentum, drawn from the primary star, which will slow the contraction of the secondary Roche lobe. Also, the matter comes from a cool disc, so it is likely to come with a reduced specific entropy, encouraging the secondary star to contract. Because of this latter effect, and because some mass may be lost to the system rather than re-accreted by the secondary, the period of the CV may shorten, following the Roche lobe relation
\begin{equation}
P_{\rm orb} \propto \left(\frac{R^{3}_{2}}{M_{2}}\right)^{1/2}
\; , 
\end{equation}
where $R_{2}$ is the radius of the secondary star. However the evolution in $P_{\rm orb}$ should be slowed as compared to the standard picture. Binaries are therefore expected to spend longer at short periods than would otherwise be expected in the weak propeller case. 

It would be useful to perform more extensive parameter space explorations and to compare the importance of magnetic and accretion torques in different configurations. A full population synthesis including binary evolution could then be used to asses the precise effect of this weak propeller model on the CV period distribution. To achieve the most realistic results, an improved model for the magnetic field would also be desirable. 

\section*{ACKNOWLEDGMENTS}
\label{sec:ack}
MRT acknowledges a PPARC postdoctoral fellowship held at the University of St Andrew between 2002 and 2005. Theoretical astrophysics research at Leicester is supported by a PPARC rolling grant. The computations reported here were performed using the UK Astrophysical Fluids Facility (UKAFF). The authors thank the organisers of The Astrophysics of Cataclysmic Variables and Related Objects in 2004, held at Strasbourg, where this work was initiated. We also thank Roland Speith for helpful discussions and Daniel Rolfe for help with visualisation. The authors are grateful to the referee, Yoji Osaki, for helpful comments.
% Star means no section number (or full width of page for figure}

\end{document}